\documentstyle[12pt]{article}

\input{tcilatex}

\begin{document}

\title{{\LARGE A Classical Analog of Quantum Search}{\bf {\Large \thanks{%
Research of the first author was partly supported by NSA\ \&\ ARO under
contract no. DAAG55-98-C-0040.}}}}
\author{Lov K. Grover, Anirvan M. Sengupta \\
{\it \{lkgrover, anirvan\}@bell-labs.com} \\
Bell Laboratories, Lucent Technologies, \\
600-700 Mountain Avenue, Murray Hill NJ 07974}
\date{}
\maketitle

\begin{abstract}
Quantum search is a quantum mechanical technique for searching $N$
possibilities in only $\sqrt{N}$ steps. We show that the algorithm can be
described as a resonance phenomenon. A similar algorithm applies in a purely
classical setting when there are $N$ oscillators, one of which is of a
different resonant frequency. We could identify which one this is by
measuring the oscillation frequency of each oscillator, a procedure that
would take about $N$ cycles. We show, how by coupling the oscillators
together in a very simple way, it is possible to identify the different one
in only $\sqrt{N}$ cycles.
\end{abstract}

\baselineskip=14pt

\newpage \pagebreak

\section{Introduction}

A\ single quantum oscillator has multiple modes of oscillation. For example
a spin $1/2$ particle in a magnetic field has two modes of oscillation and
is referred to as a {\it qubit}. Furthermore, it is in general in both of
these simultaneously. This is in contrast to a classical oscillator which
just has a single mode of oscillation. Therefore if we need $N$ modes of
oscillation, we will need at least $N$ classical oscillators, while if we
use quantum oscillators, each of which has two modes of oscillation
(qubits), we could do with just $\log _{2}N$ oscillators.

Quantum computing algorithms, such as quantum search, make use of the fact
that a quantum system is simultaneously in multiple states to carry out
certain computations in parallel in the same hardware. To implement the
actual quantum search algorithm one needs a quantum mechanical system where
one can carry out certain elementary quantum mechanical operations in a
controlled way, it is {\it not} possible to implement the algorithm on
classical hardware. Yet, in this paper we show that a very similar algorithm
works in a classical system. The difference is that in a classical system
the hardware required is proportional to $N$ whereas in the quantum system,
the hardware is only proportional to $\log _{2}N.$

The algorithm of this paper is of interest, both in its own right as a
classical algorithm and also for the insight it provides into quantum
computing. For example, it is well established that the quantum search
algorithm, which can search $N$\ possibilities in only $\sqrt{N}$ steps, is
the best possible algorithm for exhaustive searching. Yet there is no simple
argument as to why this is the best algorithm or why the algorithm should
need $\sqrt{N}$ steps. This paper gives an elementary argument as to why it
needs $\sqrt{N}$ cycles to identify the different pendulum.

\section{Background}

Any quantum mechanical transformation is a rotation of the state-vector in $%
N $\ dimensional complex Hilbert space ($N$ is the number of states).
Therefore any quantum mechanical algorithm too, is a rotation of the
state-vector in $N$\ dimensional complex Hilbert space. The quantum search
algorithm \cite{grover96} is a special case since it is a rotation in a
carefully defined two dimensional vector space. This fact was first noticed
by Farhi and Gutmann who used it to develop the following variant of the
search algorithm \cite{farhi gutmann}.

Consider an $N$\ state system, whose Hamiltonian is known to be $\left|
w\right\rangle \left\langle w\right| $. $w$ is known to be a basis state,
the problem is to find out which one this is. We are allowed to add on any
additional term to the Hamiltonian (provided this does not depend on $w$)
and let the system evolve in any way we choose. The question is as to how
rapidly can we identify $w$?

Any obvious technique will need $O(N)$ time. For example, if we examine each
state separately by coupling it to an auxiliary state, it will take $O(1)$
time to examine each state and thus $O(N)$ time in all. However, by using an
analogy with the quantum search algorithm, it is possible to devise a scheme
to identify $w$ that requires only $O(\sqrt{N})$ time.

The idea is to first add an additional term of $\frac{1}{N}(|1>+\cdots
+|N>)(<1|+\cdots +<N|)$ to the given Hamiltonian. Then start the system from
the superposition $\frac{1}{\sqrt{N}}(|1>+\cdots +|N>)$, let it evolve for a
time $O(\sqrt{N})$ and finally carry out an observation. In the following
two paragraphs we show that with a high probability the state observed will
be $\left| w\right\rangle $. This technique is similar to the search
algorithm in that it consists of a rotation of the state vector in a
two-dimensional vector space defined by $\left| w\right\rangle $ and $\frac{1%
}{\sqrt{N}}(|1>+\cdots +|N>)$.

To simplify notation, assume that $w$ is the first of the $N\;$states, i.e. $%
w=1$. The total Hamiltonian then becomes:

\begin{equation}
H=\frac{1}{N}(|1>+\cdots +|N>)(<1|+\cdots +<N|)+|1><1|.
\end{equation}

\noindent Writing this in the subspace spanned by $|1>$ and $|B>\equiv \frac{%
1}{\sqrt{N}}\sum_{j=2}^{N}|j>$, and leaving out terms of order $\frac{1}{N}$%
, the above Hamiltonian becomes: 
\begin{equation}
H\approx (|1><1|+|B><B|)+\frac{1}{\sqrt{N}}(|1><B|+|B><1|).
\end{equation}%
Thus the quantum dynamics of the system is essentially that of two {\it %
degenerate} levels with mixing amplitude of $O(1/\sqrt{N})$. The initial
state $1/\sqrt{N}\sum_{j=1}^{N}|j>\approx |B>$ ``rotates'' to $|1>$ in a
time inversely related to the mixing matrix element. Since this element is $%
O(1/\sqrt{N})$, the time taken by this search algorithm is $O(\sqrt{N})$.

\ The discrete quantum search algorithm is very similar. The main difference
is that instead of having the Hamiltonian be constant throughout, it is
adjusted so that the item specific portion acts separately from the mixing
portion, i.e. there are alternate steps of $\left| w\right\rangle
\left\langle w\right| $ and $\frac{1}{N}(|1>+\cdots +|N>)(<1|+\cdots +<N|)$.
This perspective is described in \cite{sch}. Thus, at the heart of the
search algorithm, is a resonance phenomenon. In the following sections we
discuss a classical analogue of the same phenomenon involving coupled
oscillators. Variants of the quantum search algorithm have previously been
proposed with classical waves \cite{scully}, \cite{Lloyd}, \cite{kwiat}. We
present the algorithm in a conceptually different way as a resonance
phenomenon. Using this interpretation, it can be implemented in a way that
is very different from the original search algorithm.

\section{Classical Analogy}

The analysis and results of the following two sections hold for any system
of classical oscillators, either mechanical or electrical. However, for
concreteness we consider the oscillators to be pendulums.

Consider the following problem. We are given $N$\ pendulums - one of which
is slightly shorter than the rest. The problem is to identify which one this
is. We could measure the frequency of each one of these separately and thus
identify the shorter one. This would take $O(N)$\ cycles. Instead, in the
next section, we show that by carefully coupling them together and letting
them oscillate for $O(\sqrt{N})$ cycles, a substantial portion of the energy
can be transferred to the shorter pendulum whose amplitude becomes very
high. This is accomplished by a resonance phenomenon very similar to that in
quantum search. Using this it is possible to identify the different pendulum
as briefly described in section 5.

\section{$N$ coupled pendulums}

We show that by suspending the $N$\ pendulums from a bigger pendulum (figure
1) and adjusting the masses and lengths of the bigger pendulum
appropriately, it is possible to achieve a coupling similar to that of the $%
N $\ states in the quantum search algorithm (section 2). As in section 2,
let us make the first pendulum special while the rest of the $\left(
N-1\right) $ of them are identical.

\bigskip \bigskip \bigskip 

\begin{center}
Figure 1 - {\it N}{\bf \ pendulums are suspended from a single pendulum.}
\end{center}

The Lagrangian of the system of figure 1 is given by 
\begin{equation}
L=\frac{1}{2}[M\dot{X}^{2}-KX^{2}+\frac{1}{N}(m_{1}\dot{x_{1}}%
^{2}-k_{1}(x_{1}-X)^{2})+\frac{1}{N}\sum_{j=2}^{N}(m\dot{x_{j}}%
^{2}-k(x_{j}-X)^{2})];\,\;K\equiv (M+\frac{m}{N})\frac{g}{L},k_{j}\equiv
m_{j}\frac{g}{l_{j}}  \label{Lagrangian}
\end{equation}%
where $X$ is the displacement of the support pendulum, $x_{j}$ is the
displacement of the $j^{th}$ pendulum hanging from the support; $M,L$ are
the mass and the length of the support pendulum, $\frac{m_{1}}{N},l_{1}$ are
the mass and length of the first pendulum and $\frac{m}{N},l$ are the mass
and the length of each of the other pendulums ($g$ is the acceleration due
to gravity). It was probably simpler to keep the Lagrangian of (\ref%
{Lagrangian}) in terms of the $m^{\prime }s,$ $l^{\prime }s$ and $g.$
However, as mentioned before, the framework of this paper applies to any
system of oscillators, electrical or mechanical. In order to be able to
quickly translate the results to other applications, we express the
Lagrangian in (\ref{Lagrangian}) in a more general notation in terms of the
stiffnesses ($k\prime s$).

Now we change variables so that we consider the center of mass mode $\bar{x}$
of pendulums $2,\ldots ,N$, and other modes of excitation of the same
pendulums orthogonal to the center of mass mode which we denote by:\ \ $%
y_{l},$ $l=1,..,\left( N-2\right) $. In terms of these variables, the
Lagrangian may be written as: 
\begin{equation}
L=\frac{1}{2}[M\dot{X}^{2}-KX^{2}+\frac{1}{N}(m_{1}\dot{x_{1}}%
^{2}-k_{1}(x_{1}-X)^{2})+(1-\frac{1}{N})(m\dot{\bar{x}}^{2}-k(\bar{x}%
-X)^{2})+\frac{1}{N}\sum_{l=1}^{N-2}(m\dot{y_{l}}^{2}-ky_{l}^{2})].
\label{transformed}
\end{equation}%
Note that the $y$'s decouple from the rest of the variables. If we consider
an initial condition where each $y$ is zero, they will stay zero. Hence we
can omit these variables and concentrate on the three crucial ones: $X,x_{1},%
\bar{x}$. Defining $\xi \equiv \frac{1}{\sqrt{N}}x_{1}$, and ignoring some
irrelevant $O(\frac{1}{N})$ terms, the reduced Lagrangian (without the $y$%
's) may be written as: 
\begin{equation}
L_{red}\approx \frac{1}{2}[M\dot{X}^{2}-KX^{2}+m_{1}\dot{\xi}^{2}-k_{1}(\xi -%
\frac{1}{\sqrt{N}}X)^{2}+m\dot{\bar{x}}^{2}-k(\bar{x}-X)^{2}].  \label{L_red}
\end{equation}%
$\qquad $

The Lagrangian, $L_{red},$ represents two strongly coupled degrees of
freedom, namely $X$ and $\bar{x}$, and a variable $\xi $, that is weakly
coupled to others. We can first solve the $X,\bar{x}$ system. This will give
rise to two modes with frequencies which we denote by $\omega _{1}$ and $%
\omega _{2}$. The natural frequency of the $\xi $ degree of freedom that
corresponds to the special pendulum is $\omega =\sqrt{\frac{k_{1}}{m_{1}}}$
(ignoring the $O(\frac{1}{\sqrt{N}})$ coupling $\xi $ has with the other
modes). If $\omega $ is arranged to be very close to either $\omega _{1}$ or 
$\omega _{2}$, we will have a resonant transfer of energy between the two
weakly coupled systems. The number of cycles required for significant
transfer of energy to the special pendulum varies inversely with the
coupling and will be $O(\sqrt{N})$.\footnote{%
Clearly, when the deviation of the length of the pendulum approaches zero,
there should be no energy transfer to this pendulum. Yet the previous
analysis seems to suggest that the time will be $O(\sqrt{N})\;$cycles
irrespective of the deviation. The reason for this becomes clear by
examining the frequency diagram of figure 2 when the deviation between $%
\omega $ and $\overline{\omega }$\ becomes zero. Then whatever value we
choose for $\omega _{c}$, will result in an order 1 difference between $%
\omega $ and $\omega _{2},$ i.e. we will never be able to satisfy the
resonance condition.}

\bigskip 

\bigskip \bigskip \bigskip 

Figure 2 - {\bf The center of mass mode (}$\overline{\omega })${\bf \ and
the coupling mode (}$\omega _{c})${\bf \ interact to produce two new modes (}%
$\omega _{1}\,\&$ $\,\omega _{2})${\bf , one of these (}$\omega _{2})$ {\bf %
is resonantly coupled to the oscillation mode of the different pendulum (}$%
\omega ${\bf ) with an }$O(1/\sqrt{N})${\bf \ coupling}.

\qquad 

We next analyze the three-mode system defined by the reduced Lagrangian (\ref%
{L_red}) by writing its equations of motion. The equations of motion can be
written in matrix form as follows: 
\begin{equation}
\hat{M}\ddot{\vec{X}}(t)=-\hat{K}\vec{X}(t),  \label{mx_dot_dot}
\end{equation}

\noindent \noindent where the displacement vector $\vec{X}$, the mass matrix 
$\hat{M},$ and the stiffness matrix $\hat{K},$ are defined as
follows:\noindent $\ \ \ \ \vec{X}(t)\equiv \left( 
\begin{array}{c}
X(t) \\ 
\bar{x}(t) \\ 
\xi (t)%
\end{array}%
\right) $ ,\ \ \ $\hat{M}\equiv \left( 
\begin{array}{ccc}
M & 0 & 0 \\ 
0 & m & 0 \\ 
0 & 0 & m_{1}%
\end{array}%
\right) $ ,$\ \ \ \ \ \hat{K}\equiv \left( 
\begin{array}{ccc}
K+k+\frac{k_{1}}{N} & -k & -\frac{k_{1}}{\sqrt{N}} \\ 
-k & k & 0 \\ 
-\frac{k_{1}}{\sqrt{N}} & 0 & k_{1}%
\end{array}%
\right) .$

Solving (\ref{mx_dot_dot}) by assuming a solution where the time dependence
of each component of $\vec{X}(t)$ is given by $e^{i\rho t}$. It follows
after some straightforward analysis that $\rho ^{2}$ is given by the
eigenvalues of the following matrix, $\Lambda $: 
\begin{equation}
\Lambda \equiv \hat{M}^{-\frac{1}{2}}\hat{K}\hat{M}^{-\frac{1}{2}}=\left( 
\begin{array}{ccc}
\omega _{c}^{2} & -\lambda & -\frac{k_{1}}{\sqrt{NMm_{1}}} \\ 
-\lambda & \bar{\omega}^{2} & 0 \\ 
-\frac{k_{1}}{\sqrt{NMm_{1}}} & 0 & \omega ^{2}%
\end{array}%
\right) \noindent .  \label{lambda}
\end{equation}

Here $\omega _{c}^{2}\equiv \frac{1}{M}\left( K+k+\frac{k_{1}}{N}\right) $ ($%
\omega _{c}$ corresponds to the frequency of the {\it coupling} degree of
freedom, i.e. the frequency of the large pendulum), $\ \bar{\omega}%
^{2}\equiv \frac{k}{m}$ ($\bar{\omega}$\ is the frequency of the center of
mass mode), $\omega ^{2}\equiv \frac{k_{1}}{m_{1}}$($\omega $\ is the
frequency of the deviating pendulum), $\lambda \equiv \frac{k}{\sqrt{Mm}}$($%
\lambda $ is the coupling between the large pendulum and the center of mass
mode).\noindent

Inspecting the matrix $\Lambda $ makes it clear that the first two modes are
strongly coupled, whereas the first mode is only weakly coupled to the third
mode by a term of order$\frac{1}{\sqrt{N}}$. We can thus change basis so
that the $\left( 1,2\right) $ block is diagonalized. The corresponding
frequencies are given by the eigenvalues of the $\left( 1,2\right) $ block: 
\begin{equation}
\omega _{1,2}^{2}=\frac{1}{2}\left[ \omega _{c}^{2}+\bar{\omega}^{2}\pm 
\sqrt{(\omega _{c}^{2}-\bar{\omega}^{2})^{2}+4\lambda ^{2}}\right] .
\end{equation}%
In the rotated basis, each of the first two modes will have $O(\frac{1}{%
\sqrt{N}})$ coupling with the third mode, the matrix $\Lambda $ gets
transformed into a matrix $\tilde{\Lambda}$ of the following form:

\begin{equation}
\tilde{\Lambda}\equiv \left( 
\begin{array}{ccc}
\omega _{1}^{2} & 0 & -\frac{\alpha }{\sqrt{N}} \\ 
0 & \omega _{2}^{2} & -\frac{\beta }{\sqrt{N}} \\ 
-\frac{\alpha }{\sqrt{N}} & -\frac{\beta }{\sqrt{N}} & \omega ^{2}%
\end{array}
\right) ;\,\,(\alpha ,\beta \,\,are\,\,of\,\,order\,1).  \label{alpha beta}
\end{equation}

We start this system by giving a push to the large support pendulum,
delivering order 1 energy. This energy will first be in the ($1,2)$
subsystem. However, under the condition of resonance, in $O(\sqrt{N})$
cycles, the special pendulum will swing with an amplitude of order 1. All
the other $\left( N-1\right) $ identical pendulums would move in lock step;
their total energy would be order 1, but individual pendulums will have
energy of $O(1/N)$, and their amplitudes would be $O(\frac{1}{\sqrt{N}})$.

It must be noted that precise information about the different item is
required in order to satisfy the resonance condition - we would have to know
precisely how much longer or shorter this pendulum was as compared to the
remaining pendulums. This would determine the value of $M$\ and $L$ (the
mass and length of the support pendulum from which the rest of the pendulums
are suspended).

\section{The algorithm}

As described above, we have a means for transferring a large portion of the
energy from the support pendulum into an aberrant pendulum, assuming we have
precise information about the length of this pendulum but do not know which
this is. This procedure can be used to identify which pendulum this is (as
in the quantum search algorithm). In order to better define the problem, it
is important to list some of the associated constraints.

\subsection{Rules of the game}

\begin{enumerate}
\item The system is started by giving a single push to the support pendulum.

\item We can redesign parameters and observe the motion of a constant number
of pendulums.

\item Observations can only be resolved with a finite precision that is
independent of $N.$
\end{enumerate}

These constraints are meant to reflect realistic limitations on the system.
Also, these constraints are what make the problem interesting. For example,
if we could observe the system with arbitrary precision, then we could
deduce the presence of the short pendulum by observing the motion of any
pendulum in a constant number of cycles, even without any resonance.
However, this demands a precision of $O(1/N)$.

\subsection{Algorithm}

The following procedure ascertains whether or not there is a special
pendulum in the set that is connected to the support pendulum. Once we have
a procedure for identifying the presence (or absence) of a desired item in a
specified set, it is possible to identify precisely which one this. This is
a standard technique in computer science and is accomplished by $\log _{2}N$
repetitions of the identification procedure in a binary search fashion. The
procedure is as follows.

\begin{quotation}
Select any one of the pendulums and shorten its length so that it is of the
same length as the short pendulum (assuming it is not already a short
pendulum). It is assumed that we know the length of the short pendulums. Set
the system in motion by giving a push to the support. Observe the cyclic
variation in amplitude of the shortened pendulum for $O(\sqrt{N})$ cycles.
\end{quotation}

In case the set of pendulums connected to the support already had a short
pendulum then, including the one shortened in step (1), it will have two
short pendulums. If it did not originally have a short pendulum, then it
will have just one short pendulum. An analysis similar to the previous
section shows that the resonant coupling transfers a large fraction of the
energy to and from the short pendulums with a periodicity of $O(\sqrt{N/\tau 
})$ cycles, where $\tau $ is the number of short pendulums. Thus there will
be a difference of a factor of $\sqrt{2}$ in the periodicity, depending on
whether there are 1 or 2 short pendulums. This periodicity is inferred from
the variations in amplitude in step (3).

\section{Why does it take $O(\protect\sqrt{N})$\ cycles?}

The quantum search algorithm has been rigorously proved to be the best
possible algorithm for exhaustive search, i.e. no other algorithm can carry
out an exhaustive search of $N$\ items in fewer than $O(\sqrt{N})$ steps.
The proof for this is complicated and based on subtle properties of unitary
transformations \cite{zalka}. Fortunately, in the classical analog, there is
a simple argument as to why it needs $O(\sqrt{N})$ cycles to transfer the
energy to the desired pendulum.

As described in section 4, the oscillation mode of the single pendulum is
resonantly coupled to one of the two modes arising out of interaction of the
center of mass mode (which has a mass $O(N)$\ times that of the single
pendulum) with the mode of the coupling pendulum (which too has a mass $O(N)$%
\ times that of the single pendulum). Therefore the modes that arise out of
this interaction also behave as oscillators with a mass $O(N)$\ times that
of the single pendulum.

The question is as to how rapidly can we transfer energy from a pendulum of
mass $O(N)$\ to that of a pendulum with a mass of order 1 through a resonant
coupling. Assume both pendulums to have an energy of order 1. Then the
amplitude of the larger pendulum is $O(1/\sqrt{N})$ times that of the
smaller pendulum. Since they have the same frequencies, the peak velocity of
the larger pendulum is also $O(1/\sqrt{N})$ times that of the smaller
pendulum.

Consider an elastic collision between a sphere of mass of $N$, traveling
with a velocity of $O(1/\sqrt{N}),$ with another sphere of unit mass
traveling with a velocity less than 1. As shown in figure 3, in the center
of mass frame, the larger sphere is almost stationary and the smaller sphere
bounces off the larger sphere. The speed of the smaller sphere stays
unaltered and the velocity changes sign (in order to conserve kinetic
energy). Translating back to the original frame, we see that the magnitude
of the velocity of the smaller sphere has increased by $2/\sqrt{N}.$
Therefore it will take $O(\sqrt{N})$ such collisions for the velocity of the
smaller sphere to be made to rise from 0 to 1 (or equivalently to transfer
an energy of order 1). The resonant coupling repeatedly brings back the
small sphere to interact with the larger sphere thus arranging the necessary
interactions which produce the same effects on the velocity as the
collisions.

\bigskip 

\bigskip 

\bigskip \bigskip \bigskip 

Figure 3 - {\bf When a sphere of unit mass moving with unit velocity
collides with a larger sphere of mass equal to }${\it N}${\bf \ that is
moving with a velocity of }$1/\sqrt{N},${\bf \ the magnitude of the velocity
of the smaller sphere can change by at most }$2/\sqrt{N}.$

\section{Applications \&\ Extensions}

\subsection{Counting}

Estimating the number of occurrences is an important problem in statistics
and computer science. One of the first extensions of the original quantum
search algorithm was to the problem of counting where too it gave a
square-root advantage over the best possible classical algorithm \cite%
{brassard}. As might be anticipated, our classical analog too gives a
square-root advantage over the standard estimation technique.

We are given $N$ pendulums, a small fraction of them (say $\epsilon $) are
shorter than the rest. The problem is to estimate $\epsilon .$ The standard
sampling technique is to pick a certain number of pendulums at random and
measure their oscillation frequency. Since the probability of getting a
shorter pendulum in each sample is $\epsilon ,$ it will take about $\frac{1}{%
\epsilon }$ samples before we get a single occurrence of a shorter pendulum.
Since it takes $O(1)\;$cycles to estimate the oscillation frequency of a
pendulum, it will take $O\left( \frac{1}{\epsilon }\right) $ cycles to be
able to derive any reasonable estimate of $\epsilon $. On the other hand, by
extending the technique of the previous section, it is possible to estimate $%
\epsilon $ in only $O\left( \frac{1}{\sqrt{\epsilon }}\right) $ cycles.

The approach is to suspend all $N$ pendulums from a single pendulum as in
section 4 thus coupling them. Now, as before, a resonant coupling is
designed between the shorter pendulums and the rest of the system. The
strength of this coupling is $O(\sqrt{\epsilon })$. This causes energy to
flow back and forth from the shorter pendulums with a periodicity of $%
O\left( \frac{1}{\sqrt{\epsilon }}\right) $ cycles. As in section (), we
design the first pendulum to be a short pendulum. By following its amplitude
for $O\left( \frac{1}{\sqrt{\epsilon }}\right) $ cycles, we will observe a
cyclic variation. The length of this cycle will immediately identify $%
\epsilon .$

\subsection{Mechanical applications}

Consider an application where we need to transfer energy to one of $N$
subsystems. This can be accomplished by coupling the subsystems as described
in this paper and making a slight perturbation to the subsystem into which
we want the energy to flow into. After $O(\sqrt{N})$ cycles, a large
fraction of the energy will flow into the selected subsystem. Alternatively,
if we want to transfer energy from one subsystem to another, this can be
similarly accomplished by a two-step process. First, make a perturbation to
the subsystem from which the energy is coming. If the system is now allowed
to oscillate for $O(\sqrt{N})$ cycles, the energy transfers into the support
pendulum. Now, if the perturbation is removed from the source pendulum and
made in the destination pendulum, the energy will flow from the support into
the destination pendulum. By proper design it is possible to accomplish a
lossless transfer of energy from one to another pendulum. This type of
scheme would be especially useful in an application where we need the
flexibility of transferring energy to any one of $N$ components with minimal
changes in hardware: a mechanical router.

\subsection{Quantum Mechanical Applications}

In quantum mechanical settings there are several applications where various
modes of oscillation are coupled through the {\it center of mass mode}. For
example, consider $N$ atoms coupled resonantly to a photon mode in an
optical cavity \cite{optical cavity (qdots)}. The atoms\ are trapped in the
cavity by some kind of electromagnetic fields. The photon mode plays the
role of the support pendulum through which the particles are coupled.
Consider the basis state \TEXTsymbol{\vert}$i\rangle $ to be the state where
the photon excitation is localized on the $i^{th}$ atom. Due to the coupling
there is a certain amplitude for the excitation to transfer to another atom.
Since the atoms are close together in the cavity this amplitude is the same
between any two atoms. Therefore the Hamiltonian is of the form: $%
a\sum_{i}|i\rangle \langle i|+$ $b\sum_{i,j}|i\rangle \langle j|.$This is
exactly the kind of Hamiltonian that motivated our analysis (section 2). A
similar analysis applies in the case of an ion-trap \cite{ion trap} or in
the case of Josephson junctions \cite{josephson junction} coupled through a
mutual inductance.

Our approach would carry over to the situations described above. An
important problem in an ion-trap is to transfer excitations from one ion to
another. This is the quantum analog of the problem of section 7.2. It needs
further research to decide the pros and cons of our method as compared to
the Cirac-Zoller method \cite{ion trap}.

\end{document}